\documentclass[
    ,final            
  ]
  {aipproc}

\layoutstyle{8x11single}

\usepackage{amsmath,amssymb}
\usepackage{array}
\usepackage{slashed}
\usepackage{multirow}

\newcommand{\met}{\rlap{\,\,/}E_T}

\newcommand{\genericT}{\ensuremath{T}}

\newcommand{\mptvec}{\slashed{\vec{p}}_\genericT}

\newcommand{\beq}{\begin{equation}}
\newcommand{\eeq}{\end{equation}}

\newcommand{\bea}{\begin{eqnarray}}
\newcommand{\eea}{\end{eqnarray}}

\begin{document}

\title{Measuring Properties of Dark Matter at the LHC}



\author{Kyoungchul Kong}{
  address={Department of Physics and Astronomy, University of Kansas, Lawrence, KS 66045, USA}
}

\begin{abstract}
We review various theoretical methods for measuring dark matter properties at the Large Hadron Collider.
\end{abstract}

\maketitle


\section{Introduction}
\label{sec:intro}

A compelling solution to the dark matter problem requires synergistic progress along many lines of inquiry. Especially the diversity of possible dark matter candidates requires a balanced program based on four pillars: direct detection experiments that look for dark matter interacting in the laboratory, indirect detection experiments that connect lab signals to dark matter in our own and other galaxies, collider experiments that elucidate the particle properties of dark matter, and astrophysical probes sensitive to non-gravitational interactions of dark matter such as dark matter densities in the centers of galaxies and cooling of stars.

At colliders, it is generally believed that missing energy signatures offer the best bet for discovering new physics beyond the Standard Model. 
This belief is reinforced by the dark matter puzzle.  
However, at hadron colliders the total energy and longitudinal momentum of the event are unknown.
Therefore, the production of any invisible particles can only be inferred from an imbalance in the total {\em transverse} momentum. 
The measured {\em missing} transverse momentum, $\mptvec$, then gives the sum of the transverse momenta of all invisible particles in the event.
Unfortunately, $\mptvec$ is the only measured quantity directly related to the invisible particles. Without any further 
model-dependent assumptions, it is in general very difficult, if not impossible, to make any definitive statements about the nature and properties of the missing particles. 

In this article, we review some of various proposed kinematic methods for measuring properties of dark matter at the LHC.
The main advantage of the kinematic approaches is that they 
make very few assumptions about the details of the underlying physics model 
(gauge groups, spins etc). 
This means that they can provide rather robust information,
and act as the first step towards understanding the underlying theory. 
Often the kinematic method is considered merely as a method to measure masses of particles. 
This is a completely wrong prejudice. The kinematic method is basically parameter estimation and discovery techniques.
Finding sensible variables buys more than just mass measurements, {\it e.g.,} signal sensitivity and background rejection.
There are many types of technique depending on different levels of outcome.
Some require few assumptions ($\met$, $M_{\rm eff}$, $H_T$, $\sqrt{\hat{s}}_{min}$, $\cdots$) 
while some require many (polynomial constraints, cross section method, max likelihood/matrix element method $\cdots$).
The former is relatively easy to implement and robust but lead to somewhat vague conclusions. 
The latter is hard to implement and fragile but provides specific conclusions. 
Therefore the best interpretation would require the balance of benefits. 
For a given topology/hypothesis, one must impose some interpretation, and design the variable to suit the interpretation. 
In what follows we will discuss proposals for determining dark matter properties such as mass, spin and stabilizing mechanism.

\section{Masses}
\label{sec:masses}

There are several questions regarding determination of masses 
(see \cite{Barr:2010zj,Barr:2011xt} for detailed review and more references.).
First of all, without any {\it apparent} missing particles, it is relatively easy task since 
one can reconstruct whole system of particles.
Although there still might be some combinatorial issues with many particles, in principle, 
the mass can be determined from invariant mass.
A non-trivial question is mass determination in the presence of {\it one} missing particle.
One assumes that missing transverse momentum mainly results from a missing particle.
Occasionally it is possible to reconstruct masses even in this case.
Examples include $W$ decay to a lepton plus a neutrino and semi-leptonic decay of $t\bar{t}$ system.
The former example benefits from properties of transverse mass while the latter 
uses invariant masses of $W$ and $t$.

A problematic case is mass determination in the presence of two (or more) missing particles.
In fact, many theories beyond the Standard Model are grouped into this case, 
which makes the task important and a high priority.
Those massive missing particles are potential candidates for dark matter particles.
So essentially we are looking for methods to weigh dark matter particles.
During the last decade or so, 
there have been several attempts to answer this question and 
some well known methods are the following.

\begin{itemize}

\item {\bf I. Endpoint methods.} These rely on the kinematic endpoints 
\cite{Hinchliffe:1996iu,Bachacou:1999zb,Hinchliffe:1999zc,%
Allanach:2000kt,Gjelsten:2004ki,Gjelsten:2005aw}
or shapes \cite{Miller:2005zp,Burns:2008cp}
of various invariant mass distributions constructed out of the 
visible (SM) decay products in the cascade chain.

\item {\bf II. Polynomial methods.} Here one attempts exact event reconstruction 
using the measured momenta of the SM particles and the measured
missing transverse momentum
\cite{Kawagoe:2004rz,Cheng:2007xv,Nojiri:2008ir,Cheng:2008mg}.

\item {\bf III. $M_{T2}$ methods.} These methods explore the 
transverse invariant mass variable $M_{T2}$
originally proposed in \cite{Lester:1999tx} and later used and developed in
\cite{Barr:2003rg,Lester:2007fq,Tovey:2008ui,Cho:2007qv,Nojiri:2008vq}.
Recently it was shown that under certain circumstances,
the endpoint of the $M_{T2}$ distribution, when considered 
as a function of the unknown test mass of the lightest new particle,
exhibits a kink whose location is given by the true masses of daughter and mother particles 
\cite{Cho:2007qv,Gripaios:2007is,Barr:2007hy,Burns:2008va}.

\item {\bf IV. Hybrid methods.} One could also combine two or more of these techniques
into a hybrid method, e.g.~a mixed polynomial and endpoint method \cite{Nojiri:2007pq},
a mixed $M_{T2}$ and endpoint method \cite{Ross:2007rm,Barr:2008ba},
or a mixed $M_{T2}$ and polynomial method \cite{Cho:2008tj,Cheng:2008hk}. 

\item {\bf V. Global variables and others.} 
More recent ideas include a new global and inclusive variable \cite{Konar:2008ei,Konar:2010ma}, 
energy peaks \cite{Agashe:2012fs,Agashe:2013eba}, 
cusp structure \cite{Han:2012nr,Han:2012nm}, and singularity in multi-dimensional phase space \cite{Kim:2009si}. 
See Refs. \cite{Barr:2010zj,Barr:2011xt} for a complete list of references. 

\end{itemize}

Each of these methods has advantages and disadvantages, depending on which process is considered.
For instance, invariant mass method (endpoint method) uses a decay of one side only and 
does not usually consider the decay of the other side. 
Therefore one can try to optimize discovery potential or measurement 
by properly choosing the other side of decays.
Also it does not require pair production of the same mother particle.
Longer cascade decays are better since they provide more constraints.
On the other hand, the polynomial method uses both sides of decay and, in fact, 
requires them to be identical and the length ($n$) of the decay chains long enough 
to get sufficient constraints. If not, adding the next event helps to 
increase constraints, assuming two events have the same topology.
Due to long cascade decay, often the corresponding branching fraction 
is found to be small and one may have severe combinatorial issues. 
Nevertheless, this method can result in {\it the exact} momenta of missing particles, 
if a solution exist.

\begin{figure}[t]
\centerline{
\begin{tabular}{cc}
\includegraphics[height=0.34\linewidth]{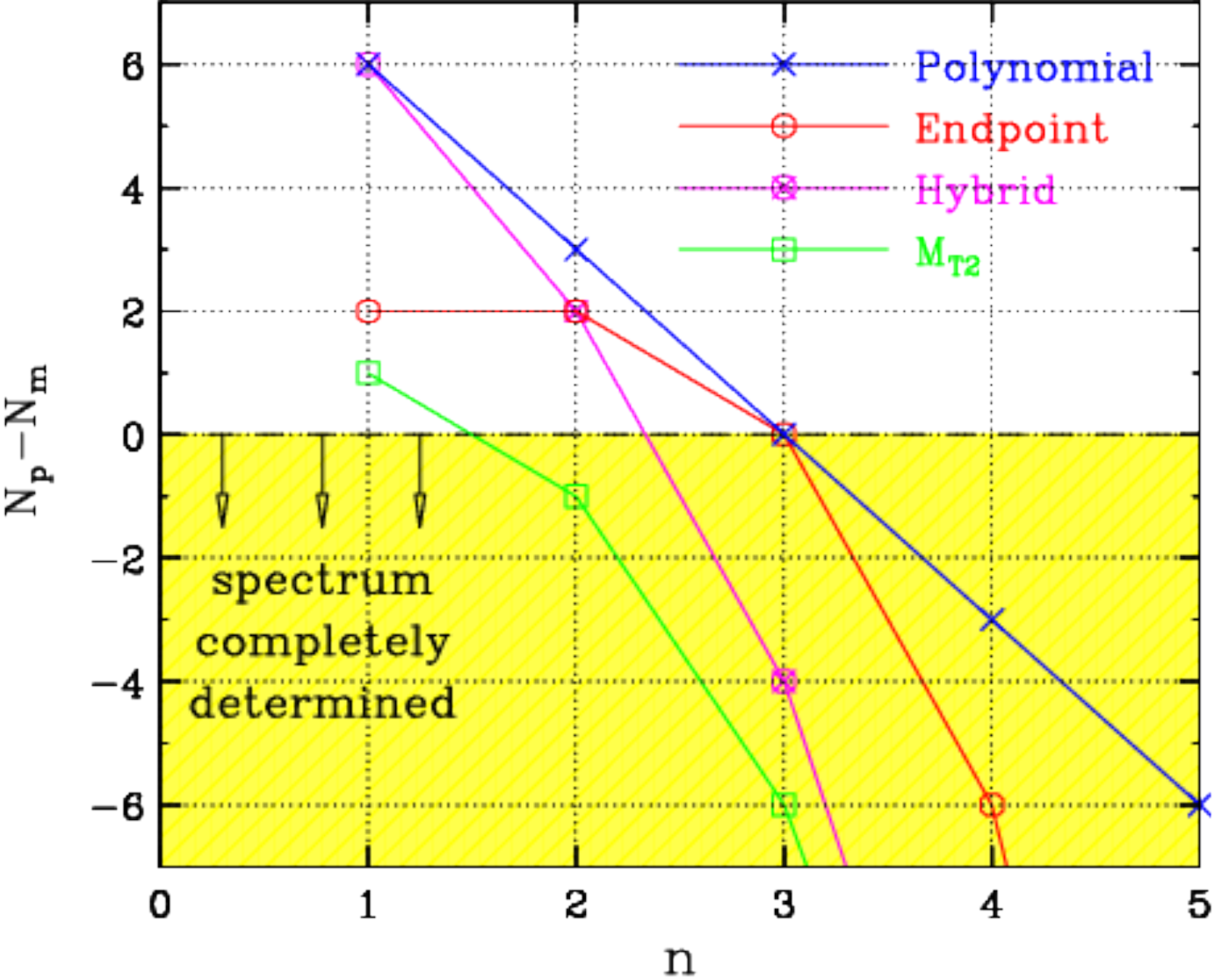}
\hspace{0.2cm}
\includegraphics[width=0.53\linewidth]{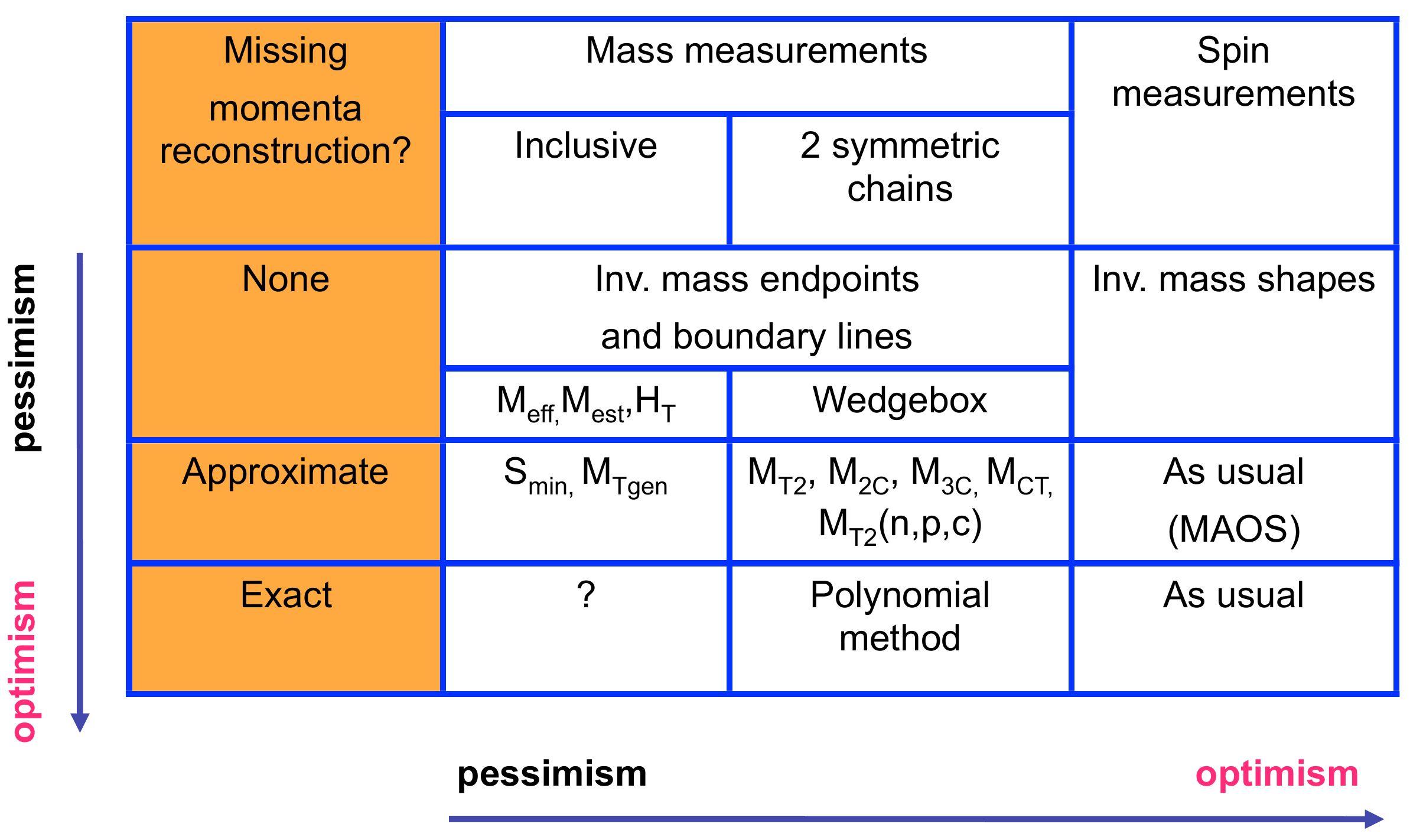}  
  &
\end{tabular}
}
\caption{\sl 
(left) The dependence of the number of undetermined parameters 
($N_p-N_m$) as a function of the number $n$ of intermediate heavy resonances in
the decay chains of the same mother, for 
various mass determination methods: $M_{T2}$ method (green, open squares),
endpoint method (red, open circles), polynomial method for $N_{ev}=2$
(blue, $\times$ symbols), or a hybrid method which is a combination
of the latter two methods (magenta, $\otimes$ symbols). Within the 
yellow-shaded region the number of unknowns $N_p$ does not exceed the 
number of measurements $N_m$ for the corresponding method, and the 
mass spectrum can be completely determined. Taken from Ref. \cite{Burns:2008va}. 
(right) Various kinematic methods for mass and spin determination.}
\label{fig:param}
\end{figure}

The standard (traditional) definition of $M_{T2}$ requires the same mothers and the same daughters, 
although intermediate particles between the mother and the daughter can be different.
The advantage of this method is that it is useful to extract a mass relation 
between mother and daughter, or in principle, both masses at once \cite{Konar:2009wn}, 
in the case of short decay chain.
Left panel in Fig.~\ref{fig:param} reveals that the endpoint method cannot
succeed unless $n\ge 3$, where $n$ is the number of two-body decays. 
This conclusion has already been confirmed by numerous studies of various low-energy 
SUSY models, where one considers a decay chain of sufficient length:
$n=3$ as in the squark decay, or 
$n=4$ as for a gluino chain \cite{Gjelsten:2005aw}.
On the other hand, if $n=1$ or $n=2$, 
with this method we are unable to pin down {\em all}
of the new particle masses, even as a matter of principle.
These are exactly the cases where the additional information
from mass measurements at future lepton colliders
has been seen as extremely useful.
 We see that for $n\le 3$, the
$M_{T2}$ method is by far the most powerful, and more importantly, 
it is the only method which is able to handle the problematic 
case of $n=2$! 

$M_{T2}$ methods are especially useful for shorter decays ($ n \leq 2$). 
End point of $M_{T2}$ distribution as a function of a trial mass exhibits an interesting kink structure, 
that is useful for determination of masses of mother and daughter particles.
The first type of kink arises when considering more than one visible particle, whose invariant mass varies event by event.
The second type shows up when system of two mother particles are boosted due to initial state radiation (ISR).
The stronger ISR returns more pronounced kink structure.
The last one is from decays of particles, which are heavier than dark matter candidate.

which allows simultaneous determination of all particles involved in the decay chain.

The right panel of Fig. \ref{fig:param} shows various kinematic methods for measuring masses and spins, depending on level of their assumptions.
Many of these existing mass measurement variables proposed for hadron colliders are 
far more closely related to each other than is widely appreciated, and 
indeed can all be viewed as a common mass bound specialized for a variety of purposes \cite{Barr:2011xt}.

\section{Spins and Couplings}
\label{sec:spins}

\begin{figure}[t]
\centerline{
\includegraphics[width=0.6\linewidth]{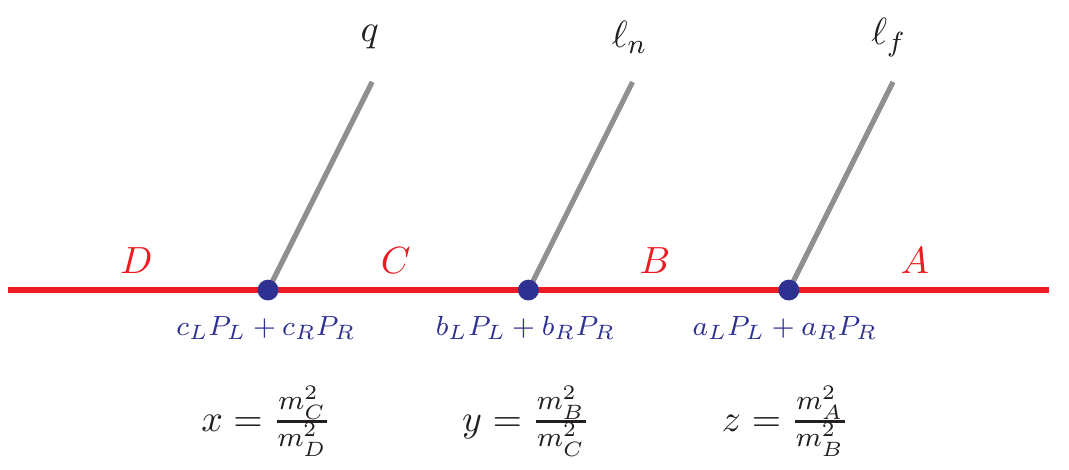}  
 \caption{
The typical UED or SUSY cascade decay chain under 
consideration.
At each vertex, the most general fermion couplings are assumed
(see Ref.~\cite{Burns:2008cp} for the exact definitions).}
\label{fig:ABCD}}
\end{figure}

{\bf \underline{Invariant mass method}}: 
It is now very well known that the kinematic endpoints 
in the invariant mass distributions give relations between masses of particle 
involved in the cascade decay and their shapes share spin information \cite{Smillie:2005ar,Athanasiou:2006ef,Datta:2005zs,Kong:2006pi}.
Recently in Ref. \cite{Burns:2008cp}, it was discussed that the invariant mass distributions not only have 
spin information but also have information of some combination of couplings and mixing angles, 
which make it more difficult to extract spin information out of shape of invariant distributions. 
This can be best illustrated with an example. 
Consider the three-step decay chain shown in Fig.~\ref{fig:ABCD},
which is typical in both Universal Extra Dimensions (UED) and SUSY models.
The measured visible decay products are a jet $j$ 
and two opposite sign leptons $\ell^+$ and $\ell^-$,
while the end product $A$ is invisible in the detector.
Given this limited amount of information, in principle 
there are 6 possible spin configurations for the 
heavy partners $D$, $C$, $B$ and $A$: $SFSF$, $FSFS$, $FSFV$,
$FVFS$, $FVFV$, and $SFVF$, where $S$ stands for a spin-0 scalar,
$F$ stands for a spin-$\frac{1}{2}$ fermion, and 
$V$ stands for a spin-1 vector particle.
The main goal of the invariant mass analysis 
is to discriminate among these 6 possibilities, and in particular
between $SFSF$ (SUSY) and $FVFV$ (UED).

Unfortunately the invariant mass distributions are also
affected by a number of additional factors, which have 
nothing to do with spins, such as: 
the chirality of the couplings at each vertex \cite{Burns:2008cp};
the fraction of events $f$ in which the cascade 
is initiated by a particle $D$ rather than its antiparticle $\bar{D}$; and finally, 
the mass splittings among the heavy partners \cite{Smillie:2005ar}.
Therefore, in order to do a pure and model-independent 
spin measurement, one has to somehow eliminate the effect 
of those three extraneous factors.
Most likely the masses of $A$, $B$, $C$ and $D$ can be 
determined ahead of time, 
by measuring the kinematic endpoints of various invariant mass distributions 
\cite{Hinchliffe:1996iu,Allanach:2000kt,Gjelsten:2004ki,Gjelsten:2005aw},
or through a sufficient number of transverse mass measurements \cite{Lester:1999tx,Burns:2008va}. 
But we are still left with a complete lack of knowledge regarding the 
coupling chiralities and particle fraction $f$.
In spite of this residual ambiguity, the spins can 
nevertheless be determined, at least as a matter of principle 
\cite{Burns:2008cp}. To this end, one should not make any
a priori assumptions and instead consider 
the most general fermion couplings at each vertex in Fig.~\ref{fig:ABCD}
and any allowed value for the parameter $f$.
Then, the invariant mass distributions should be used to make 
separate independent measurements of the spins, on one hand, and 
of the couplings and $f$ fraction, on the other.

Given the three visible particles from the decay chain of Fig.~\ref{fig:ABCD},
one can form three well-defined two-particle invariant mass distributions: 
one dilepton ($\ell^+\ell^-$), and two jet-lepton ($j\ell^+$ and $j\ell^-$) distributions.
For the purposes of the spin analysis, it is actually more convenient to consider 
the sum and the difference of the two jet-lepton distributions.
The shapes of the resulting invariant mass distributions are given
schematically by the following formulas \cite{Burns:2008cp}:
\begin{eqnarray}
\left( \frac{dN}{dm^2_{\ell\ell}}\right)_S &=&
F_{S;\delta}^{(\ell\ell)}(m^2_{\ell\ell})
+ \alpha\, F_{S;\alpha}^{(\ell\ell)}(m^2_{\ell\ell}) 
\label{L+-} \\ [2mm]
\left( \frac{dN}{dm^2_{j\ell^+}}\right)_S + 
\left( \frac{dN}{dm^2_{j\ell^-}}\right)_S &=&
F_{S;\delta}^{(j\ell)}(m_{j\ell}^2)
+ \alpha\, F_{S;\alpha}^{(j\ell)}(m_{j\ell}^2)  
\label{S+-} \\ [2mm]
\left( \frac{dN}{dm^2_{j\ell^+}}\right)_S -
\left( \frac{dN}{dm^2_{j\ell^-}}\right)_S &=&
 \beta \, F_{S;\beta }^{(j\ell)}(m_{j\ell}^2) 
+\gamma\, F_{S;\gamma}^{(j\ell)}(m_{j\ell}^2)\ ,
\label{D+-}
\end{eqnarray}
where the functions $F$, given explicitly in \cite{Burns:2008cp},
are known functions of the masses of particles $A$, $B$, $C$ and $D$. 
As indicated by the 
index $S$, there is a separate set of $F$ functions for each
spin configuration: 
%
$S=\{SFSF, FSFS,  FSFV,FVFS, FVFV, SFVF \} \, .$
%
Thus the functions $F$ contain the 
pure spin information. On the other hand, the coefficients
$\alpha$, $\beta$ and $\gamma$ encode 
all of the residual model dependence, namely
the effect of the coupling chiralities and 
particle-antiparticle fraction $f$. Since 
the coefficients $\alpha$, $\beta$ and $\gamma$ are a priori 
unknown, they will need to be determined from experiment, 
by fitting the predicted shapes (\ref{L+-}-\ref{D+-}) to the data.

The general solution for the couplings can be written in terms of the measured parameters $\alpha$, $\beta$ and $\gamma$, as
\begin{eqnarray}
\begin{array}{ll}
    |a_L| = \frac{1}{\sqrt{2}} \left( 1 \pm \frac{1}{\beta }\, \sqrt{ \alpha\beta\gamma } \right)^{\frac{1}{2}},  
& |a_R| = \frac{1}{\sqrt{2}} \left( 1 \mp \frac{1}{\beta }\, \sqrt{ \alpha\beta\gamma } \right)^{\frac{1}{2}}, \\
    |b_L| = \frac{1}{\sqrt{2}} \left( 1 \pm \frac{1}{\gamma}\, \sqrt{ \alpha\beta\gamma } \right)^{\frac{1}{2}},  
&  |b_R| = \frac{1}{\sqrt{2}} \left( 1 \mp \frac{1}{\gamma}\, \sqrt{ \alpha\beta\gamma } \right)^{\frac{1}{2}}, \\ 
     |c_L| = \frac{1}{\sqrt{2}} \left( 1 \pm \frac{1}{f-\bar{f}}\, \frac{1}{\alpha}\, \sqrt{ \alpha\beta\gamma } \right)^{\frac{1}{2}},  
&   |c_R| = \frac{1}{\sqrt{2}} \left( 1 \mp \frac{1}{f-\bar{f}}\, \frac{1}{\alpha}\, \sqrt{ \alpha\beta\gamma } \right)^{\frac{1}{2}},  
\end{array}
\end{eqnarray}
where the appearance of the $\pm$ sign is due to the two-fold ambiguity in the definition of chirality, {\it i.e.,} 
we can only measure the chirality of the three different vertices in Fig.~\ref{fig:ABCD} only {\em relative} to each other.
Note that there are conditions on $\alpha$, $\beta$ and $\gamma$.
The product $\alpha\beta\gamma$ is always non-negative. Furthermore, from their definitions 
it also follows that $|\alpha\beta | \le |\gamma|$,
$|\beta \gamma| \le |\alpha|$ and $|\gamma\alpha| \le |\beta |$.
Therefore all square roots in above equations are well behaved and never yield any imaginary solutions. 
It is interesting to note the dependence on the particle-antiparticle fraction $f$. 
We see that for any given measurement of $\alpha$, $\beta$ and $\gamma$, the effective couplings 
$|a_L|$, $|a_R|$, $|b_L|$ and $|b_R|$ associated with the particle A and particle B vertices of Fig.~\ref{fig:ABCD} can be 
uniquely determined, up to the two-fold $L\leftrightarrow R$ ambiguity.
Although we do not know the exact value of $f$, consistency of 
above equations restricts the allowed values of $f$ to be in the range
\begin{equation}
0\le f \le \frac{1}{2}\left( 1-\sqrt{\frac{\beta\gamma}{\alpha}} \right) \quad {\rm or} \quad
\frac{1}{2}\left( 1+\sqrt{\frac{\beta\gamma}{\alpha}} \right)  \le f \le 1\ .
\label{frange}
\end{equation}
The allowed range for $f$ splits into two separate intervals. 
At a $pp$ collider like the LHC, in general we expect $f>\frac{1}{2}$,
so we should select the higher $f$ range in eq.~(\ref{frange}), while 
the lower $f$ range in eq.~(\ref{frange}) would be relevant for
a hypothetical $\bar{p}\bar{p}$ collider (``anti-LHC''):
\begin{eqnarray}
{\rm      LHC}\,(pp)             &:&\quad \frac{1}{2}\left( 1+\sqrt{\frac{\beta\gamma}{\alpha}} \right)  \le f \le 1\ , \label{frangeLHC}\\
{\rm anti-LHC}\,(\bar{p}\bar{p})&:&\quad 0\le f \le \frac{1}{2}\left( 1-\sqrt{\frac{\beta\gamma}{\alpha}} \right)\ . \label{frangeantiLHC}
\end{eqnarray}
While eq.~(\ref{frangeLHC}) is not a real measurement of the value of $f$
at the LHC, it nevertheless contains very important information.
For example, if the measured values of $\alpha$, $\beta$ and $\gamma$ 
happen to be such that $|\beta\gamma| \approx |\alpha|$, then $f$ becomes
very severely constrained, and the restriction (\ref{frangeLHC}) 
by itself is sufficient to yield a measurement of the value of $f$: $f\approx 1$.  

All these methods are developed analytically and all functions are derived in literature \cite{Burns:2008cp}. 
However it is very important to illustrate the methods including appropriate detector simulation and backgrounds. 
Also decays of vector particles are missing in the spin configuration. \\

{\bf \underline{Spin measurements from production cross-sections}}: 
The spin of the new particles can also be inferred from the 
threshold behavior of their production cross-section
\cite{Battaglia:2005zf}.
For an $s$-channel diagram mediated by a gauge boson, 
the pair production cross-section for a spin-0 particle 
behaves like $\sigma \sim \beta^3$ 
while the cross-section for a spin-$\frac{1}{2}$ 
particle behaves as $\sigma \sim \beta$, 
where $\beta = \sqrt{1 - \frac{4m^2}{s}}$, and 
$\sqrt{s}$ is the total center-of-mass energy, while $m$ is the 
mass of the new particle. At lepton colliders the threshold behavior 
can be easily studied by varying the beam energy and
measuring the corresponding total cross-section, without any need 
for reconstructing the kinematics of the missing particles.
In contrast, at hadron colliders the initial state partons cannot be controlled, so in order to apply this method, one has to  
fully reconstruct the final state, which is rather difficult
when there are two or more missing particles.

The total production cross-section may also be used as an indicator
of spin \cite{Kane:2008kw}. For example, 
the total cross-sections of the fermion KK modes in UED are 
5-10 times larger than the corresponding cross-sections for
scalar superpartners of the same mass. However, the measurement
of the total cross-section necessarily involves additional 
model-dependent assumptions regarding the branching fractions, 
the production mechanism, etc. \\

{\bf \underline{Spin measurements from angular distributions}}:
Perhaps the most direct indication of the spin of the new particles is 
provided by the azimuthal angular distribution at production 
\cite{Battaglia:2005zf}.
Assuming a pair production through an $s$-channel gauge boson,
the angular distribution for a spin-0 particle is $\sim (1 - \cos^2\theta)$,
where $\theta$ is the azimuthal production angle in the center-of-mass frame.
In contrast, the distribution for a spin-$\frac{1}{2}$ particle 
is $\sim (1 + \cos^2\theta)$.
Unfortunately, reconstructing the angle $\theta$ 
generally requires a good knowledge of the momentum 
of the missing particles, which is only possible at a lepton collider.
Applying similar ideas at the LHC, one finds 
that typically quite large luminosities are needed \cite{Barr:2005dz}. \\

{\bf \underline{Spin measurements from quantum interference}}:
When a particle is involved in both the production and the decay, its spin $s$
can also be inferred from the angle $\phi$ between the production and decay planes
\cite{Buckley:2008pp}.
The cross section can be written as 
\begin{equation}
\frac{d\sigma}{d \phi} = a_0 + a_1 \cos\phi + a_2 \cos 2\phi + \cdots + a_{2s} \cos 2s \phi\ .
\end{equation}
By measuring the coefficient $a_{2s}$ of the highest $\cos$ mode, 
one can in principle extract the spin $s$ of the particle.
This method is especially useful since it does not rely on the particular
production mechanism, and is equally applicable to $s$-channel and $t$-channel processes.
However, its drawback is that the $\phi$ dependence results from
integrating out all other degrees of freedom, which often leads 
to a vanishing coefficient as a result of cancellations, 
for instance, in the case of a purely vector-like coupling, or in the case of 
a $pp$ collider like the LHC.
As a result, the practical applicability of the method is rather model-dependent.

\section{Stability and Asymmetric Event Topologies}
\label{sec:two}

Tremendous amount of effort has been made to reconstruct events with dark matter candidate at the LHC in order to determine the masses of the DM, the mother particles and possibly intermediate particles in the decay chains. Most studies consider a $Z_2$-parity for the stability of dark matter. This is because the most popular models, e.g. supersymmetric (SUSY), little Higgs and extra-dimensional scenarios, all ensure the dark matter candidates remain stable by employing a $Z_2$ stabilization symmetry.
However, any discrete or continuous global symmetry can be used to stabilize dark matter. 
Furthermore, because all fundamental particles in nature are defined by how they transform under various symmetries, 
most of the popular ($Z_2$) models actually consider only one type of DM candidate. 
It is therefore critical to determine experimentally, i.e., without theoretical bias, the nature of the symmetry that stabilizes dark matter.
Distinguishing dark matter stabilization symmetries has been investigated 
utilizing multiple kinematic edges, cusps, and shape of $M_{T2}$ in Refs. \cite{Agashe:2010gt,Agashe:2010tu}, 
where especially $Z_3$ symmetry is compared to $Z_2$. It also also noticed that the cusp is generally invariant of the various spin configurations. 

\begin{figure}[t]
\centerline{
\includegraphics[width=0.45\linewidth]{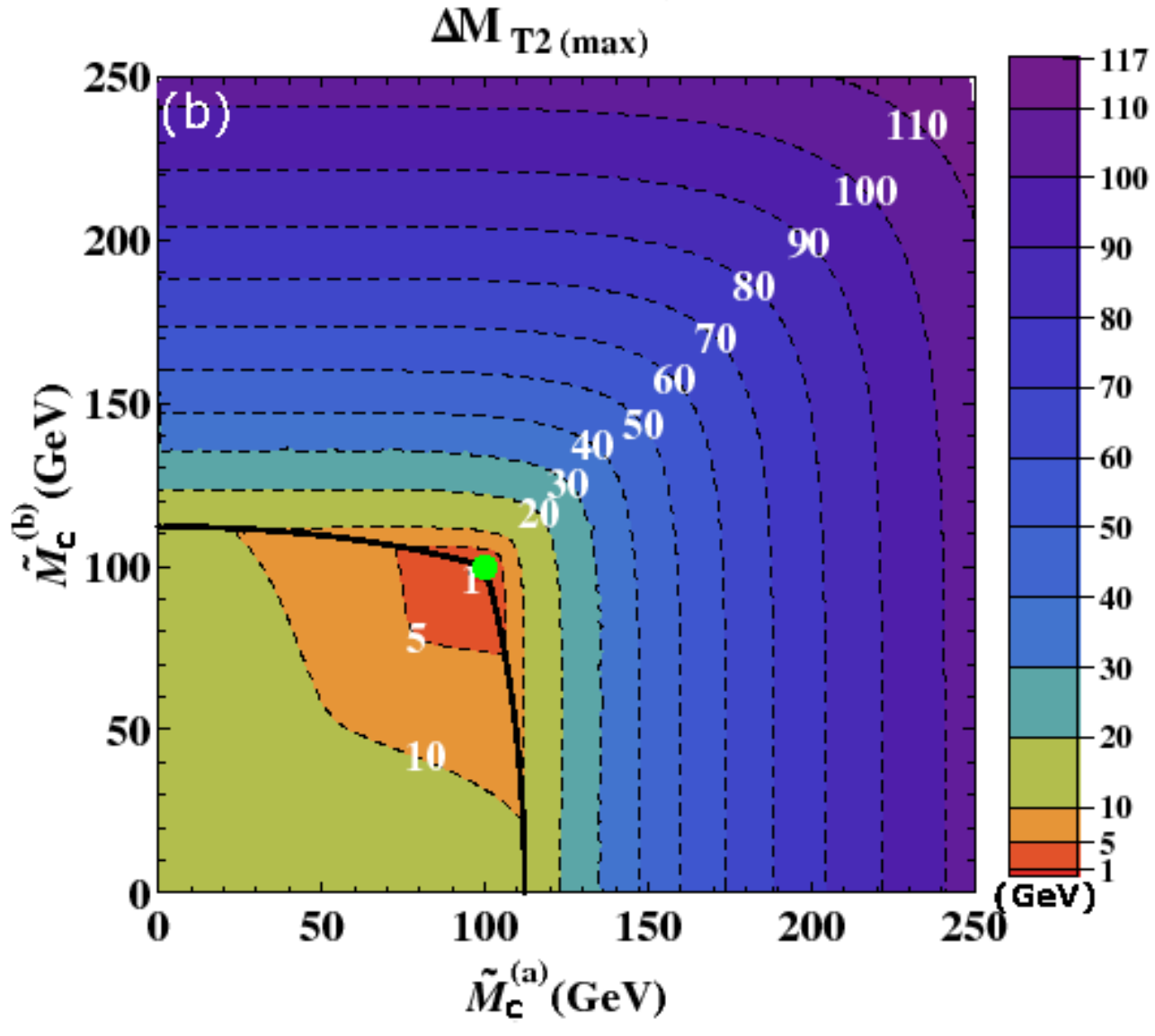}  \hspace{0.4cm}
\includegraphics[width=0.47\linewidth]{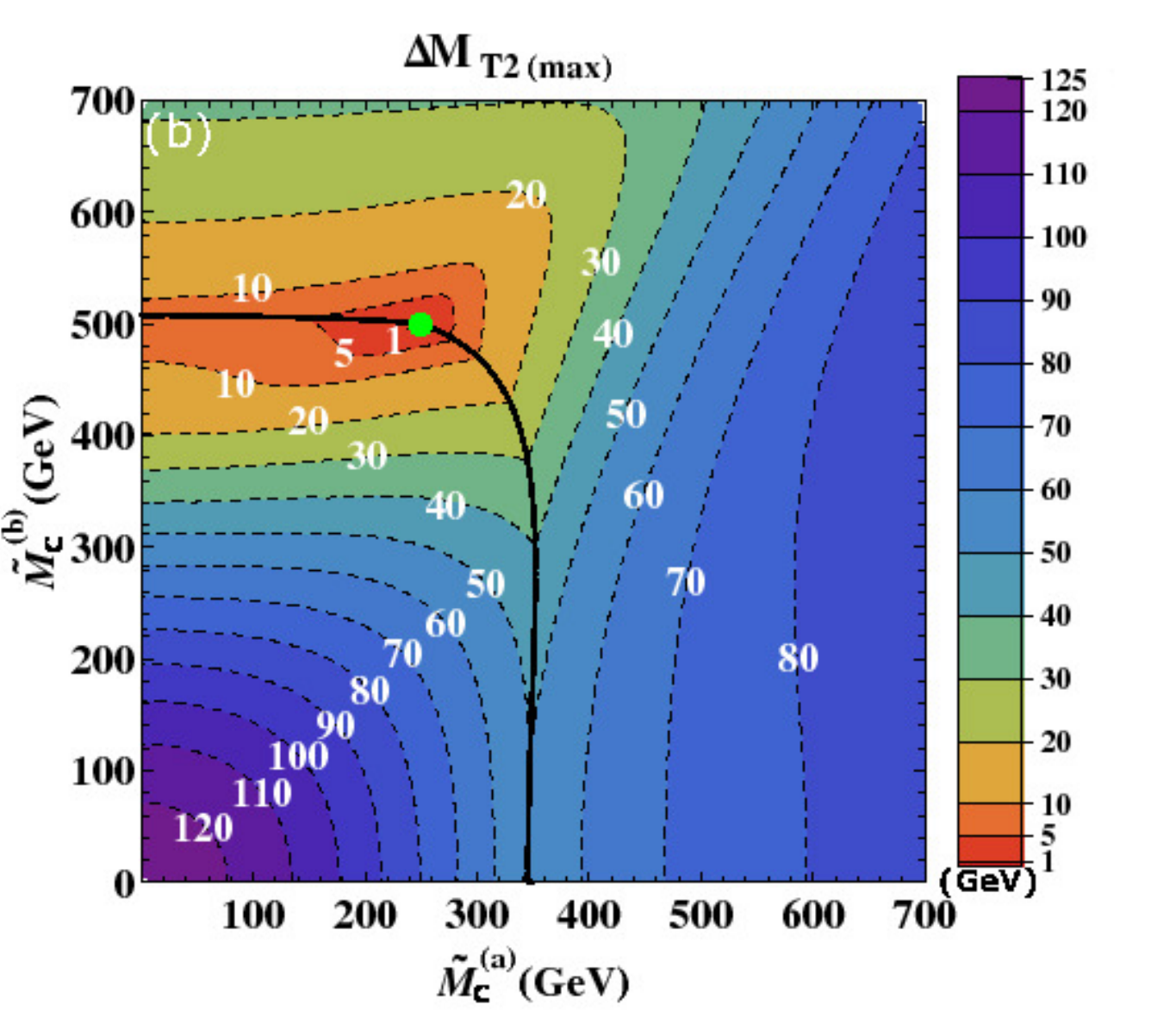}   }
\caption{ \it 
Contour plot of 
$\Delta M_{T2(max)}= M_{T2(max)}(P_{UTM}=1\ {\rm TeV}) -M_{T2(max)}(P_{UTM}=0\ {\rm TeV}) $ 
in the $(\tilde M_c^{(a)},\tilde M_c^{(b)})$ plane, for identical daughters of mass 100 GeV (in left), and 
non-identical daughters of masses (300, 500) GeV (in right).
The mother particles masses are 300 and 600 GeV, respectively.
Notice that the minimum of the $\Delta M_{T2(max)}$ function is 
now obtained at $\tilde M_c^{(a)}=\tilde M_c^{(b)}$ in the left, 
implying that the two missing particles are the same, while in the right it is not.
The solid black curve indicates the location of the $M_{T2(max)}$ ridge.  
Only the point corresponding to the true children masses (the green dot) 
satisfies the $P_{UTM}$ invariance condition $\Delta M_{T2(max)}=0$.
Taken from Ref. \cite{Konar:2009qr}.
}
\label{fig:110dlspridge}
\end{figure}

In Ref. \cite{Konar:2009qr} similar questions are asked in a different way: 
if there is more than one candidate, can we distinguish them?
One can generalize analysis to the case with two non-identical daughter particles, which makes 
the $M_{T2}$ as a function two test masses and the one dimensional relation becomes 
a surface in 3 dimensional mass space \cite{Konar:2009qr}. 
Fig. \ref{fig:110dlspridge} shows two examples: 
one for the identical daughters and the other for non-identical daughters.
The $\Delta M_{T2(max)}$ function plotted there is the difference between two $M_{T2}$'s with or without $\vec{P}_T$ (upstream transverse momentum).
By looking at this figure, one can recognize whether two missing particles are identical or not from 
the shape of contours \cite{Konar:2009qr}.
We do not know, {\it a priori}, whether missing energy originates from one particle 
or two or more. Further more, we do not know whether they are the particles of the same mass!

Possible applications of the asymmetric $M_{T2}$ idea are as follows.
(i) Invisible decays of the next-to-lightest particle. Most new physics models introduce some new massive and neutral particle which plays the role of a dark matter candidate. Often the very same models also contain other, heavier particles, which for collider purposes behave just like a dark matter candidate: they decay invisibly and result in missing energy in the detector. For example, in supersymmetry one may find an invisibly decaying sneutrino, 
$\tilde \nu_\ell \to \nu_\ell \tilde \chi_1^0$. 
(ii) Applying $M_{T2}$ to an asymmetric subsystem. One can apply the $M_{T2}$ idea even to events in which there is only one (or even no) missing particles to begin with. Such an example is $t\bar{t}$ production in the dilepton or semi-leptonic channel. In the first leg we can take $b\ell$ as our visible system and the neutrino $\nu_\ell$ as the invisible particle, while in the other leg we can treat the b-jet as the visible system and the $W$-boson as the child particle. In this case, there still should be a ridge structure revealing the true $t$, $W$ and $\nu$ masses.
(iii) Multi-component dark matter. Of course, the model may contain two (or more) different genuine dark matter particles, 
whose production in various combinations will inevitably lead at times to asymmetric event topologies.

\section{Outlook}
\label{sec:outlook}

These kinematic variables not only provide information of masses and spins but are also very useful for 
{background rejection} since we know what is expected in SM. 
They guide us where to look in search for new physics. 
Therefore they can be used for discovery as well as measurements. 
They are also used to understand combinatorial backgrounds in events.
Especially $M_{T2}$ and invariant mass methods have been considered often and their usefulness 
has been illustrated in Ref. \cite{Alwall:2009zu}. 
Consider a pair production of gluinos and their three-body decays to neutralinos giving $4 j + \met$ as the signal. 
Suppose the five hardest jets are selected. Excluding a jet, say the $i$-th jet ($i=1, \dots 5$), 
one can form $M_{T2}^{(i)}= M_{T2}$ with the rest of the 4 jets. 
Then $min ( M_{T2}^{i})$ is guaranteed to be bounded by the expected maximum endpoint and 
the minimum selects which jet should be the ISR jet. 
Statistically this method gives correct $P_{T}$ and $\eta$ distributions of an ISR jet.
This idea can be applied to the $t\bar{t}$ system to resolve a combinatorial 
issue with $b$-jets and a lepton (or two leptons). 
This greatly improves probability of finding a correct pair \cite{Baringer:2011nh}.
This observation deserves more attention and needs detailed simulation including backgrounds.

\begin{figure}[t]
\centerline{
\includegraphics[width=0.98\linewidth]{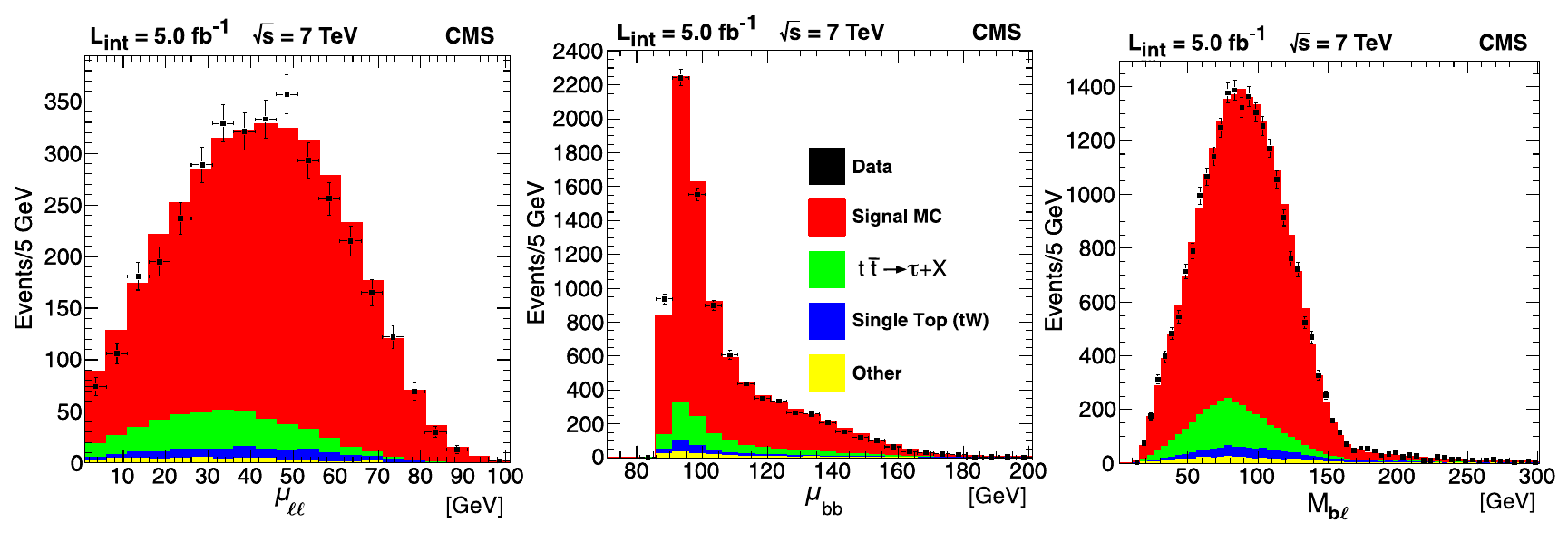}  
 \caption{
\it Distributions of the three kinematic distributions, $\mu_{\ell\ell}$, $\mu_{b\ell}$ and $M_{b\ell}$. Taken from Ref. \cite{Chatrchyan:2013boa}.
}
\label{fig:ttbar}}
\end{figure}

These methods are not only used for new physics search but also used for SM physics \cite{Chatrchyan:2013boa}. 
Recently CMS investigated a simultaneous measurement of the top-quark, $W$-boson, and neutrino masses in the dilepton final state 
from a data sample corresponding to an integrated luminosity of 5.0 fb$^{-1}$ collected at $\sqrt{s} = 7$ TeV. 
The analysis is based on endpoint determinations in kinematic distributions. 
When the neutrino and W-boson masses are constrained to their world-average values, a top-quark mass value of 
$M_t  = 173.9 \pm 0.9 {(\rm stat.)} ^{+1.7}_{-2.1} {(\rm syst.)}$ GeV is obtained. 
When such constraints are not used, the three particle masses are obtained in a simultaneous fit. 
In this unconstrained mode the study serves as a test of mass determination methods that may be used in beyond standard model physics scenarios 
where several masses in a decay chain may be unknown and undetected particles lead to under-constrained kinematics.
Fig. \ref{fig:ttbar} shows kinematic distributions of $\mu_{\ell\ell}$, $\mu_{bb}$ and $M_{b\ell}$ with data taken by CMS collaboration. 
The $\mu_{\ell\ell}$ variable, known as $M_{T2\perp}^{220}$, uses the two leptons of the $t\bar{t}$ dilepton decays, treating the neutrinos as lost child particles, and combining the $b$-jets with all other upstream momentum in the event. 
The $\mu_{bb}$ variable, known as $M_{T2\perp}^{221}$, uses the $b$-jets, and treat the $W$ bosons as lost child particles (ignoring the fact that their charged daughter leptons are in fact observable). It considers only ISR jets as generators of upstream momentum. 
Together with the invariant mass distribution, $M_{b\ell}$, $M_{T2}$ subsystem can be constructed for determination of all three masses. 

%


\begin{theacknowledgments}
We are grateful to the organizers of PPC 2013 and our attendance at PPC 2013 is supported in part by the US DOE Grant DE-FG02-12ER41809 and by the University of Kansas General Research Fund allocation 2301566. 
\end{theacknowledgments}



\bibliographystyle{aipproc}   





\end{document}